\documentclass{Interspeech2024}

\interspeechcameraready 
\usepackage{svg}
\usepackage{lipsum}  
\usepackage{url}
\usepackage{amsmath,array,graphicx}
\usepackage{kantlipsum}

\title{Improving Data Augmentation-based Cross-Speaker Style Transfer for TTS with Singing Voice, Style Filtering, and F0 Matching}

\name[affiliation={1,2}]{Leonardo B. de M. M.}{Marques}
\name[affiliation={1,2}]{Lucas H.}{Ueda}
\name[affiliation={2}]{Mário}{U. Neto}
\name[affiliation={2}]{Flávio}{O. Simões}
\name[affiliation={2}]{Fernando}{Runstein}
\name[affiliation={2}]{Bianca}{Dal Bó}
\name[affiliation={1}]{Paula D. P.} {Costa}

\address{ $^1$Universidade Estadual de Campinas (UNICAMP)\footnote{Department of Computer Engineering and Automation (DCA), School of Electrical and Computer Engineering}, $^2$CPqD, Brazil}

\email{paulad@unicamp.br, lmenezes@cpqd.com.br}


\begin{document}

\maketitle

\begin{abstract}

The goal of cross-speaker style transfer in TTS is to transfer a speech style from a source speaker with expressive data to a target speaker with only neutral data. In this context, we propose using a pre-trained singing voice conversion (SVC) model to convert the expressive data into the target speaker's voice. In the conversion process, we apply a fundamental frequency (F0) matching technique to mitigate tonal variances between speakers with significant timbral differences. A style classifier filter is proposed to select the most expressive output audios for the TTS training. Our approach is comparable to state-of-the-art with only a few minutes of neutral data from the target speaker, while other methods require hours. A perceptual assessment showed improvements brought by the SVC and the style filter in naturalness and style intensity for the styles that display more vocal effort. Also, increased speaker similarity is obtained with the proposed F0 matching algorithm.
\end{abstract}


\section{Introduction}


To build a text-to-speech (TTS) system in a certain speech style, the traditional approach is to directly record data or get samples of the target speaker in the desired speaking style. However, automatic style labeling of audio in the wild is error-prone and recording is usually expensive and demands both time and the speaker's ability to articulate in the desired style, becoming, in most situations, unfeasible~\cite{li_2022_cross-speaker}. Several attempts to tackle this issue exploit the so-called cross-speaker style transfer task. This approach consists in the transfer of stylistic speech of a speaker (source), to another speaker (target) that only has neutral data. The task is relevant to the field of expressive speech synthesis since it enables a speaker with only neutral data to speak expressively, alleviating the burden of the recording process.

Data augmentation methods have been successfully employed in low-resource scenarios~\cite{hubrechts_2021_low}. Recent works have proposed the use of data augmentation to generate synthetic expressive data for style transfer through a voice conversion (VC) model, subsequently employing these data to train a TTS model. 

The first work using data augmentation with a VC model used the CopyCat~\cite{karlapati_2020_copycat} model to convert the source utterances and a Tacotron 2~\cite{shen_2018_tacotron} TTS model to generate speech~\cite{hubrechts_2021_low}. However, the method still required some expressive data from the target speaker. A text-predicted-GST-Tacotron2~\cite{stanton_2018_predicting}, which learns to predict stylistic renderings from text, was used on an on-the-fly-based data augmentation setting that consisted of approximating target and source speakers' attention matrices and style embeddings~\cite{chung_2021_onthefly}. Due to training from scratch, more than $11$ hours of target speaker's neutral data, and at least $2$ hours from each style were required. Another work extends the first attempt by using an improved version of CopyCat with fundamental frequency conditioning~\cite{qian_2020_f0}. Yet, $10$ hours of target speaker neutral speech and $1$ hour on the given style are still required~\cite{ribeiro_2022_cross}. In~\cite{zhang_2022_cross}, the amount of stylized data is further reduced to as low as fifteen minutes by using a curriculum learning strategy. 


Only one work addressed the challenge of cross-speaker style transfer with highly expressive styles, which relies on the VC to preserve the emotion on conversion~\cite{terashima_2022_cross}. They combined the VC with pitch-shifted augmented data in a wide range of pitch dynamics. It dealt with two speaking styles in a proprietary dataset: happy and sad. However, it only used speakers of the same gender, which may have a similar timbre. It also required 1000 utterances from the target speaker's neutral speech.


This work focuses on the open problem of cross-speaker style transfer with highly expressive styles and gets inspiration from singing voice conversion (SVC).
Compared to speech, singing voice includes richer emotional information, longer continuous pronunciations, and enhanced higher frequency formants~\cite{huang_2021_multisinger}. There are four phonation modes in singing voice: pressed, breathy, flow, and modal, which arise from variations in pitch, loudness, and mainly from larynx level adjustments. They produce different voice qualities, such as rich harmonic content in pressed voice, a laxed voice with lots of airflows on the breathy mode, a higher loudness with minimum effort on the flow voice, and a more resonant voice arising from the modal mode~\cite{rouas_2016_automatic}. Hence, inferring a correlation between singing voice and stylistic speech is reasonable. For example, the higher frequency content of the pressed mode can be heard in an angry style. Thus, we propose substituting the VC on the cross-speaker style transfer pipeline with an SVC model, as we expect it to capture and preserve the expressive style better.  Our approach and contributions are summarized as follows:

\begin{figure*}[t]
  \centering
  \includegraphics[width=\textwidth]{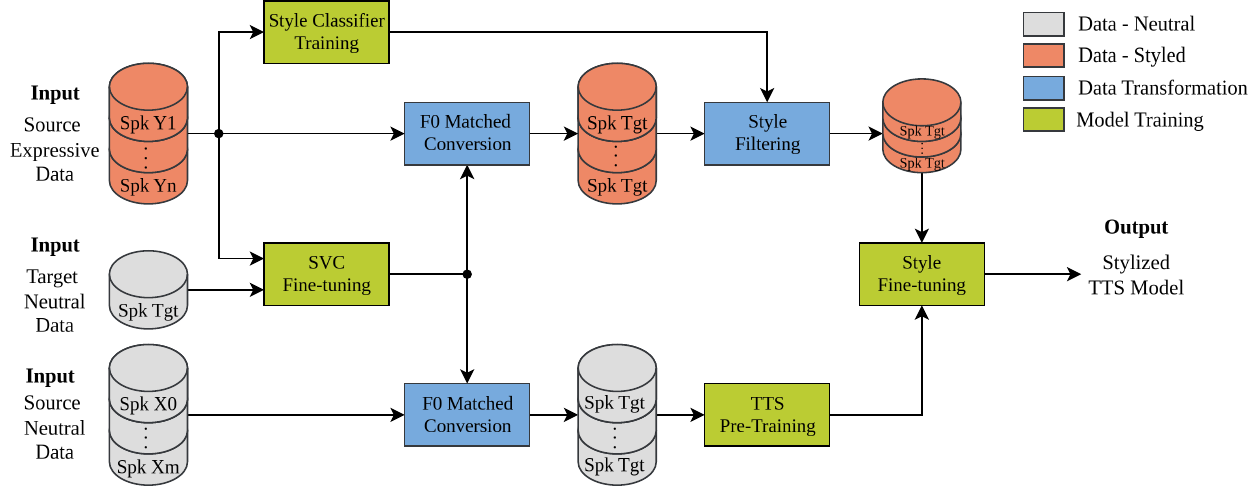}
  \caption{Overview of the method investigated for building a TTS model with target speaker's voice and source speaker's style.}
  \label{fig:method}
\end{figure*}

\begin{itemize}
    \item We proposed to augment data with an SVC model to better capture the expressiveness of the source speakers for the cross-speaker style transfer task.
    \item We designed an F0 matching technique that mitigates timbral differences between target and source speakers.
    \item A style classifier filter was employed to select the most expressive converted data to perform the style finetuning.
    \item We proposed to use transfer learning and a base neutral dataset to reduce the amount of the target speaker's neutral data, lowering this value to only a few minutes.
    \item Only open-source data and models were used. All the generated models, code, and audios are made available\footnote{Resources are available at \url{svcstytransfer.netlify.app}}. 
    \item We compared the proposed techniques with several other methods to perform cross-speaker style transfer, providing a perceptual evaluation benchmark for the current research state.
\end{itemize}

\section{Method}

\subsection{Overview}
An overview of the proposed cross-speaker style transfer pipeline based on data augmentation is shown in Figure~\ref{fig:method}. The pipeline receives three inputs: an expressive dataset, spoken by $N$ source speakers (in orange); a neutral base dataset, spoken by other $M$ source speakers; and target speaker's neutral dataset (in gray). First, to convert both source neutral and expressive datasets into target speaker's voice, a pre-trained SVC model is fine-tuned simultaneously on the target speaker's data and also on the source expressive dataset to learn the style patterns. Then, in the conversion stage, using the finetuned SVC, the source neutral and source expressive datasets are converted to the target speaker's voice using the F0 matching algorithm, in which each input source audio is transposed to match the mean fundamental frequency of the target speaker's audios. Then, a base TTS is pretrained on the neutral converted dataset. Also, a style classifier is trained with the source expressive dataset. It is used to select data from the target speaker's converted expressive dataset by choosing the audios that maintained the same style label after conversion. Finally, this filtered data is used to perform a finetuning of the base TTS, resulting in a TTS model on the target speaker's voice with the desired style.

\subsection{Singing Voice Conversion}
As our SVC model, we used So-VITS-SVC\footref{fnlabel}, a state-of-the-art, open-source, conditioned-on-F0 model. This model combines four different audio encoders that extract representations with different meanings. A pretrained timbre encoder is used to extract speaker representations, a Whisper~\cite{radford_2023_whisper} encoder is used to extract content information, a soft HuBERT~\cite{van2022comparison} model to extract prosody representation, and a CREPE~\cite{kim_2018_crepe} model to obtain the F0. These are consumed by a Flow-based decoder that generates the spectrogram. The model is also trained with a speaker classifier with a gradient reversal layer to achieve speaker disentanglement. Our SVC is pretrained simultaneously on the VCTK~\cite{veaux_2017_vctk}, to learn various timbres, and on the OpenSinger Chinese~\cite{huang_2021_multisinger} dataset to learn the conversion of richer phonation modes that occur in singing voice. Using the pretrained model enables the need for only a few minutes of target speaker's neutral speech. In the pipeline, this model was finetuned with the source expressive and the neutral target speaker datasets.

\subsection{Fundamental Frequency Matched Conversion}
The mismatch of F0 range between speakers with different vocal ranges, more prominently shown when considering different genders, can cause unrealistic converted speech. To mitigate this issue, we employed an F0 matching algorithm during the pipeline's conversion steps. For each neutral audio in all three input datasets, a mean value of F0 was computed using only the voiced segments with the Harvest estimator~\cite{morise_2017_harvest}. Then, for each speaker, a mean of the per audio F0 means was calculated, obtaining an average value of F0 in which that speaker speaks most of the time. Afterward, a distance in the interval of semitones was computed from the target speaker's average F0 value to each source speaker's. During conversion, the input F0 curve was transposed by the calculated semitonal distance between the input source speaker and target speaker's, aiming to ensure that all converted speech is in a range adequate for the target speaker's voice.

\subsection{Style Filtering}

A style filtering step was proposed to filter out the synthetic converted data that is not classified with the desired style. A style classifier was trained on the source expressive dataset until convergence. With a sufficiently large number of speakers the classifier can learn to classify each style of the dataset in a robust way regardless of the speaker's timbre. Thus, speech samples in the synthetic expressive dataset were filtered out whenever the inferred style was different from the true style. This way, using the classification results, only speech samples whose style label was kept constant after conversion were used for finetuning. A Reference Encoder~\cite{skerry_2018_towards} with a classifier linear layer on top was used as the classifier.

\subsection{Text-to-Speech}
The TTS training steps were divided into neutral pre-training and style finetuning, which generates a TTS in a specific style and speaker. The neutral pre-training, assuming the neutral source dataset had enough volumetry to train a TTS model, alleviated the requirement of a large volumetry of source expressive data and the target speaker's speech. Thus, there must be an amount of expressive source data that is just enough for the style finetuning task. We employed Fastpitch as our TTS model, with explicit duration, pitch, and energy predictors, due to its fast and high-quality TTS capability~\cite{lancucki_2021_fastpitch}.

\section{Experimental Setup}

\subsection{Datasets}
To ensure reproducibility, only open-source data was used. As our target speaker dataset, due to its well established use in TTS research, we used varying length excerpts of the single-speaker neutral style LJSpeech dataset, summing up to $5$ minutes of speech. As the neutral source dataset, we used the VCTK. This corpus was chosen not only for containing sufficient data to train a TTS model from scratch but also with the intention of highlighting the effectiveness of the F0 matched conversion, since it contains 110 different speakers with various timbres and genders, which challenges the conversion step. Finally, as the source expressive dataset, we used the English portion of the ESD dataset~\cite{zhou_2021_esd} due to its highly expressive emotions, equally shared across 10 different gender-balanced English speakers. 

\subsection{Baseline Models}

The proposed method was compared to Daft-Exprt~\cite{zaidi_2022_daft}, an open-source state-of-the-art cross-speaker prosody transfer model, designed to capture high and low-level prosodic features such as pitch, duration, and energy. Instead of using data augmentation, the authors attempt to disentangle speaker information from the prosodic through adversarial training with a gradient reversal layer. The model was trained on the full LJ Speech dataset together with the expressive dataset (ESD in our case). Two methods were considered for synthesis. First, we used the ground truth speech samples with the exact text as a reference to perform the style transfer. Secondly, prosody embeddings computed for all training samples of the ESD dataset were used as references and then the inference of a given style was performed by taking the centroid of all training prosody embeddings of that particular style~\cite{kwon_2019_effective}. We also performed some ablation studies: removal of the style filter, i.e., the whole converted ESD dataset was used on the fine-tuning, and also the latter strategy plus the replacement of SVC by an open source state-of-the-art VC model, FreeVC~\cite{lj_2023_freevc}, pretrained on the VCTK. 

\subsection{Training Setup}
All models were trained on a Quadro RTX 5000 GPU (48GB), with the hyperparameters used as default in each original implementation\footnote{So-VITS available at \url{github.com/PlayVoice/whisper-vits-svc}.\\Daft-Exprt available at \url{github.com/ubisoft/ubisoft-laforge-daft-exprt}.\\FreeVC available at \url{github.com/OlaWod/FreeVC}.\\FastPitch available at \url{github.com/AI-Unicamp/TTS}\label{fnlabel}}. The VC/SVC were finetuned for $100$ epochs, with a learning rate set to a tenth of the original. All base TTS models were trained for 600k steps from scratch, and the style fine tunings were performed for $100$k steps with a tenth of the original learning rate. The style filter was trained on the ESD data until convergence and achieved an accuracy of 84$\%$ on a validation set. When applied to the synthetic data, 571 files were selected for the final TTS fine-tuning step by the classifier for the angry style, 74 audios for the happy style; 2230 audios for the sad style; and 99 audios for the surprise style. The complete pipeline took about 6 and a half days (27 hours for the SVC finetuning, 2 hours for the style classifier training, 4 hours for each F0 matched conversion, 100 hours for the base TTS, and 20 hours for the style fine-tuning). To equally convert the generated mel-spectrograms into audio, we used the BigVGan~\cite{lee_2023_bigvgan} vocoder, a universal pretrained\footnote{BigVGan available at \url{github.com/NVIDIA/BigVGAN}.}, open-source, and state-of-the-art neural vocoder trained on the VCTK, LJSpeech, and LibriTTS datasets with a batch size of 32 for 5M steps.

\subsection{Evaluation}

A perceptual study was conducted in three subjective dimensions: naturalness, style intensity, and speaker similarity, using mean opinion scores (MOS) from $1$ to $5$. For this task, $30$ native English speakers were recruited through the Prolific\footnote{Prolific available at \url{https://prolific.com}.} platform. We used five neutral samples from a test set of the LJSpeech to evaluate the naturalness. Ground truth audios were used as a high anchor for naturalness. Style intensity was also evaluated with $3$ audio files from each of the $4$ styles present in the ESD test set, evenly distributed across speakers. Ground truth audios from the ESD were used as high anchors for style, and neutral audios from the ESD as the low anchors, all of them resynthesized. Additionaly, we evaluated speaker similarity with a test set of $2$ audios from each of the $4$ styles of the ESD, evenly distributed across speakers. The reference sample was always an LJSpeech sentence. We used samples from LJSpeech as a high anchor and samples from another male speaker of the ESD as the low anchor, all of them resynthesized. Additionally, with the aid of the state-of-the-art speaker verification model Resemblyzer~\cite{wan_2018_resemblyzer}, we evaluated the effect of the proposed F0 matching algorithm. We computed a speaker embedding for each converted samples and calculated the mean embedding for each speaker. Then, for conversions performed both with and without the F0 matching algorithm, the cosine similarity between the embedding of each converted speaker to LJ's speaker was compared with the ground truth LJ speaker embedding.

\section{Results}

\begin{table}[h]
  \caption{Naturalness MOS with 95\% confidence intervals.}
  \label{tab:naturalness}
  \centering
  \begin{tabular}{cc}
    \toprule
    \multicolumn{1}{c}{\textbf{Model}} & 
    \multicolumn{1}{c}{\textbf{MOS}} \\ 
    \midrule
    GT (High Anchor) & $4.05\pm0.18$ \\
    Daft-Exprt (Reference) & $2.01\pm0.18$ \\
    Daft-Exprt (Centroid) & $2.21\pm0.20$ \\
    VC & $3.02\pm0.20$  \\
    SVC (Ours) & $3.57\pm0.20$  \\
    \bottomrule
  \end{tabular}
\end{table}

\begin{table*}[t]
  \caption{Style Intensity Mean Opinion Scores MOS  with 95\% confidence intervals.}
  \label{tab:style}
  \centering
  \begin{tabular}{ccccc}
    \toprule
    \multicolumn{1}{c}{\textbf{}} &  \multicolumn{4}{c}{\textbf{MOS}} \\ 
    \textbf{Model} & Angry & Happy & Sad & Surprise \\
    \midrule
    GT-Res (High Anchor) & $3.85\pm0.24$ & $4.22\pm0.22$ & $4.26\pm0.23$ & $4.66\pm0.13$ \\
    Neutral-Res (Low Anchor) & $1.81\pm0.19$ & $1.86\pm0.19$ & $1.78\pm0.21$ & $1.34\pm0.12$\\
    Daft-Exprt (Reference) & $1.97\pm0.26$ & $2.43\pm0.25$ & $3.28\pm0.30$ & $2.72\pm0.25$ \\
    Daft-Exprt (Centroid) & $2.16\pm0.27$ & $2.00\pm0.18$ & $3.19\pm0.27$ & $2.40\pm0.19$ \\
    VC & $2.29\pm0.23$ & \textbf{$3.53\pm0.25$} & $2.61\pm0.24$ & $4.32\pm0.19$  \\
    SVC (Ours) & $1.94\pm0.21$ & $3.06\pm0.21$ & $2.61\pm0.24$ & $2.76\pm0.22$  \\
    SVC (Ours) w/ Filter & \textbf{$2.69\pm0.29$} & $2.00\pm0.20$ & $2.79\pm0.24$ & $3.13\pm0.26$ \\
    \bottomrule
  \end{tabular}
\end{table*}

\begin{table*}[t]
  \caption{Speaker Similarity MOS with 95\% confidence intervals.}
  \label{tab:speaker}
  \centering
  \begin{tabular}{ccccc}
    \toprule
    \multicolumn{1}{c}{\textbf{}} &  \multicolumn{4}{c}{\textbf{MOS}} \\ 
    \textbf{Model} & Angry & Happy & Sad & Surprise \\
    \midrule
    LJ-Res (High Anchor) & $3.90\pm0.30$ & $4.47\pm0.22$ & $4.28\pm0.28$ & $4.40\pm0.27$ \\
    Other-Res (Low Anchor) & $1.22\pm0.23$ & $1.13\pm0.19$ & $1.00\pm0.00$ & $1.00\pm0.00$\\
    Daft-Exprt (Reference) & $2.23\pm0.32$ & $2.55\pm0.33$ & $2.23\pm0.30$ & $2.26\pm0.32$ \\
    Daft-Exprt (Centroid) & $2.33\pm0.34$ & $2.63\pm0.33$ & $2.05\pm0.30$ & $2.67\pm0.36$ \\
    VC & $1.58\pm0.25$ & $1.48\pm0.24$ & $1.62\pm0.23$ & $1.43\pm0.18$  \\
    SVC (Ours) & $1.68\pm0.24$ & $1.71\pm0.25$ & $1.50\pm0.18$ & $1.53\pm0.23$  \\
    SVC (Ours) w/ Filter & $1.82\pm0.30$ & $1.67\pm0.28$ & $1.47\pm0.16$ & $1.43\pm0.23$ \\
    \bottomrule
  \end{tabular}
\end{table*}

\begin{figure}[t]
  \centering
  \includegraphics[width=\linewidth]{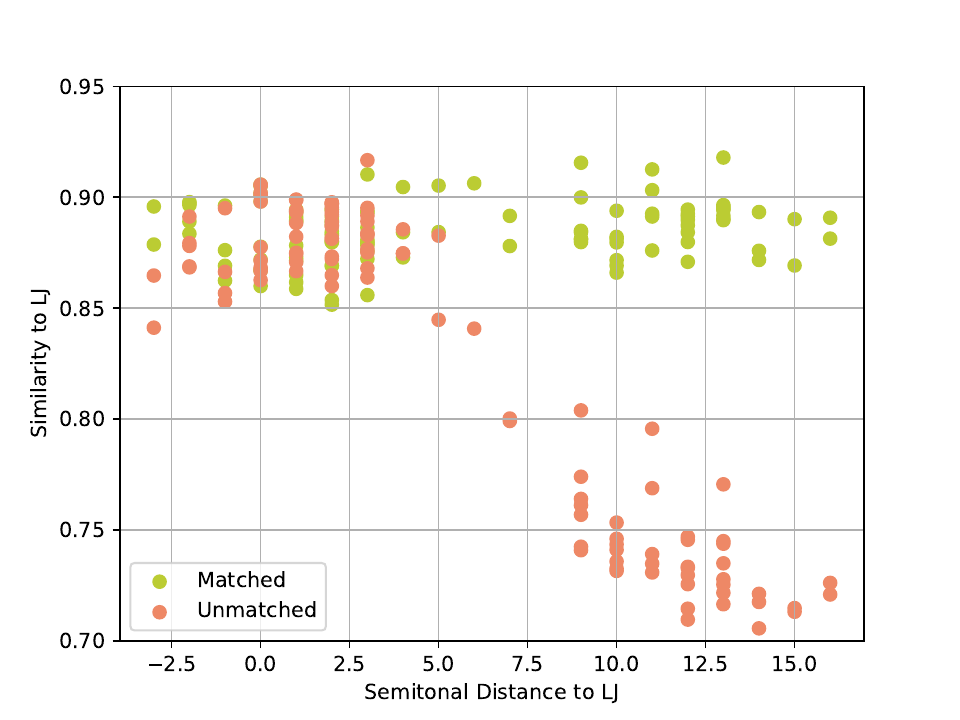}
  \caption{Similarity for each converted speaker.}
  \label{fig:spksim}
\end{figure}

Naturalness results are shown in Table~\ref{tab:naturalness}. It is seen that the SVC model improved the naturalness of the system by a great margin. Also, we found that the reference-based DaftExprt was very sensitive to the input audio, in a way that, when atempting to copy prosody exactly, unrealistic speech was generated. Style Intensity results are shown in Table~\ref{tab:style}.  For the angry style, the SVC w/ Filter model performed better by a large margin, and in all other styles, the SVC-based models obtained at least the second greater MOS. Our hypothesis is that, out of all styles, the angry was the only one that required a specific phonation mode to be correctly perceived~\cite{birkholz2015contribution}, while the others could rely primarily on prosodic parameters. Thus, with the pre-training in more diverse phonation modes, the SVC performance is highlighted on the angry style. In the Sad case, the reference-based Daft-Exprt outperformed all data augmentation-based methods, probably due to the model's ability to copy the duration aspect, which is key for this style since it is characterized by slower speaking rates. Also, we can observe the effectiveness of the style classifier filter since, for all styles but the happy, these models (SVC w/ Filter) boosted the intensity MOS when compared to the SVC without the filter. In fact, the decrease in MOS for the happy style could be attributed to the very low volumetry of the filtered happy style (77 audios), not being enough to learn its defining patterns during the fine tuning. Subjective results of the perceptual experiments for speaker similarity are shown in Table~\ref{tab:speaker}. In all three experiments our SVC-based models trained with only $5$ minutes of the target speaker's voice achieved, when not better, at least competitive results compared with the VC-based pipelines and the Daft-Exprt, which employed $24$ hours for training. Moreover, we observed that using a smaller amount of target speaker data also worked, but with a trade-off in speaker similarity. Objective speaker similarity results obtained with the Resemblyzer are shown in Figure~\ref{fig:spksim}. Each dot in the graph represents the similarity between the original LJ and a source speaker converted to LJ's voice. We plot the speakers in ascending order of their semitonal distance to the LJ. As expected, we observe a decline in speaker similarity as the semitonal difference between LJ and the speaker increases. However, when the F0 matching algorithm was applied, the similarity levels remained unaffected by the timbral difference, demonstrating the effectiveness of our technique.

\section{Conclusion}

We presented a novel approach based on an SVC model as part of a pipeline for cross-speaker style transfer. We also developed techniques to reduce the tonal mismatch between speakers with different voice qualities, to lower the amount of target speaker data needed to only a few minutes, and to enhance the expressiveness of the synthetic speech. We evaluated our pipeline against several baseline methods using subjective and objective measures of naturalness, style intensity, and speaker similarity. Our results showed that our pipeline performed better on styles with high vocal effort. We also found that the style classifier improved the style intensity MOS for all styles, but only when enough filtered data was available for fine-tuning. As future work, we plan to further decrease the target speaker data requirement to a few-shot setting.

\section{Acknowledgements}
This study is partially funded by the Coordenação de Aperfeiçoamento de Pessoal de Nivel Superior – Brasil (CAPES) – Finance Code 001, and it is supported by the BI0S - Brazilian Institute of Data Science, grant \#2020/09838-0, São Paulo Research Foundation (FAPESP). Paula D. P. Costa, Leonardo B. de M. M. Marques, and Lucas H. Ueda are also affiliated with the Dept. of Computer Engineering and Automation (DCA), Faculdade de Engenharia Elétrica e de Computação and are part of the Artificial Intelligence Lab., Recod.ai, Institute of Computing, UNICAMP.
\bibliographystyle{IEEEtran}
\bibliography{template.bib}

\end{document}